\documentclass[letterpaper]{article} 
\usepackage[submission]{aaai2026}  
\usepackage{times}  
\usepackage{helvet}  
\usepackage{courier}  
\usepackage[hyphens]{url}  
\usepackage{graphicx} 
\usepackage{natbib}  

\usepackage{caption} 
\usepackage{algorithm}
\usepackage{algorithmic}

\usepackage{hyperref}

\emergencystretch=\maxdimen
\usepackage{booktabs} 
\usepackage{makecell}   
\usepackage{tabularx}  
\usepackage{ragged2e}  
\usepackage{newfloat}
\usepackage{listings}
\DeclareCaptionStyle{ruled}{labelfont=normalfont,labelsep=colon,strut=off} 
\floatstyle{ruled}
\newfloat{listing}{tb}{lst}{}
\floatname{listing}{Listing}
\title{Small Models, Big Support: A Local LLM Framework for Educator-Centric Content Creation and Assessment with RAG and CAG}
\author{
    Zarreen Reza$^{\rm 1}$,
    Alexander Mazur$^{\rm 1}$,
    Michael T. Dugdale$^{\rm 2}$,
    Robin Ray-Chaudhuri$^{\rm 2}$
}
\affiliations{
    $^{\rm 1}$JACOBB\\
    $^{\rm 2}$John Abbott College\\
    zarreen.reza@jacobb.ai, alexander.mazur@jacobb.ai, michael.dugdale@johnabbott.qc.ca, robin.ray-chaudhuri@johnabbott.qc.ca
}

\begin{document}

\maketitle

\begin{abstract}
While Large Language Models (LLMs) are increasingly applied in student-facing educational tools, their potential to directly support educators through locally deployable and customizable solutions remains underexplored. Many existing approaches rely on proprietary, cloud-based systems that raise significant cost, privacy, and control concerns for educational institutions. To address these barriers, we introduce an end-to-end, open-source framework that empowers educators using small (3B-7B parameter), locally deployable LLMs. Our system is designed for comprehensive teacher support, including customized teaching material generation and AI-assisted assessment. The framework synergistically combines Retrieval-Augmented Generation (RAG) and Context-Augmented Generation (CAG) to produce factually accurate, pedagogically-styled content. A core feature is an interactive refinement loop, a teacher-in-the-loop mechanism that ensures educator agency and precise alignment of the final output. To enhance reliability and safety, an auxiliary verifier LLM inspects all generated content. We validate our framework through a rigorous evaluation of its content generation capabilities and report on a successful technical deployment in a college physics course, which confirms its feasibility on standard institutional hardware. Our findings demonstrate that carefully engineered, self-hosted systems built on small LLMs can provide robust, affordable, and private support for educators, achieving practical utility comparable to much larger models for targeted instructional tasks. This work presents a practical blueprint for the development of sovereign AI tools tailored to the real-world needs of educational institutions.
\end{abstract}


\section{Introduction}
\label{sec:introduction}

Large Language Models (LLMs) have shown immense potential to transform education, offering capabilities for automated tutoring, question answering, and content creation \cite{kasneci2023chatgpt}. Current state-of-the-art (SOTA) models like GPT-5 can generate high-quality learning objectives and course materials, significantly reducing the manual effort for educators \cite{sridhar2023harnessingllmscurriculardesign}. However, the practical adoption of these powerful models in many educational institutions is severely hampered by significant operational barriers. Their immense size necessitates reliance on proprietary, cloud-based APIs, leading to accumulating pay-per-token costs that are often unsustainable for resource-constrained budgets \cite{wang2024largelanguagemodelseducation, naveed2024comprehensiveoverviewlargelanguage}. Furthermore, routing sensitive educational materials and student data through third-party services raises critical concerns regarding data privacy, security, and institutional sovereignty \cite{irugalbandara2024scalingscaleupcostbenefit}, creating a major roadblock for their widespread and responsible integration.

To address these challenges, there is a growing interest in smaller, open-weight LLMs (SLMs) that can be deployed locally on institutional hardware \cite{sharir2020costtrainingnlpmodels, borzunov2023distributedinferencefinetuninglarge}. This self-hosted approach directly resolves data privacy concerns by ensuring that all sensitive information remains within the institution's secure environment. By leveraging efficient, quantized models, local deployment becomes feasible even on consumer-grade hardware without high-end GPUs, making AI-powered tools more accessible and affordable for a broader range of educational settings. This shift empowers institutions to build customized, sustainable AI solutions that align with their specific pedagogical needs and policies.

While SLMs offer a solution for deployment, their smaller size means they possess less parametric knowledge, making them more susceptible to generating factually incorrect or ``hallucinated" content. To overcome this limitation, our framework synergistically combines two powerful techniques: Retrieval-Augmented Generation (RAG) and Context-Augmented Generation (CAG). RAG grounds the LLM's output in factual sources by first retrieving relevant passages from a trusted knowledge base, such as a course textbook or curriculum documents, and providing them to the model as context \cite{li2024enhancingllmfactualaccuracy}. CAG is a broader technique that enriches the model's prompt with additional context to steer its output, such as pedagogical style guides or examples of existing exercises, ensuring the generated content aligns with the educator's specific instructional intent. Together, RAG and CAG enable smaller models to produce factually accurate, relevant, and stylistically appropriate educational content.

Recent work in AI for Education has begun to demonstrate the efficacy of this local, augmented approach. For instance, several studies have successfully used SLMs with RAG to create student-facing tools like Q\&A assistants and programming tutors \cite{yu2025can, hicke2023aitaintelligentquestionanswerteaching, ma2024integratingaitutorsprogramming}. These systems show that a carefully configured RAG pipeline can allow a small model to rival the performance of much larger models on domain-specific tasks \cite{yu2025integrating}. However, the majority of existing work focuses on student support. A comprehensive, \textit{teacher-centric} framework that supports the full instructional design workflow, from customized content and rubric creation to assessment assistance, all within a unified, private, and interactive environment remains a significant and underexplored area.

This paper introduces a novel, end-to-end open-source framework designed to fill this gap, providing educators with a powerful, private, and practical AI assistant. Our framework leverages locally deployed SLMs (3B-7B parameters) to support teacher-centric content creation and assessment. By focusing on institutional sovereignty, data privacy, and small-model efficiency, our work contributes a practical direction for deploying AI systems that serve the needs of educators directly. The main contributions of this paper are:
\begin{itemize}
    \item A novel, end-to-end open-source framework leveraging small, locally deployable LLMs for comprehensive teacher support, including customized content generation, rubric creation, and AI-assisted grading.
    \item The synergistic application of Retrieval and Context Augmented Generation (RAG/CAG) for producing factually accurate educational content that is aligned with specific pedagogical styles.
    \item An interactive, teacher-in-the-loop refinement process that ensures educator agency and control over the final output, a crucial feature for the efficacy of smaller models.
    \item A jailbreak and hallucination mitigation strategy using an auxiliary verifier LLM to enhance system safety and reliability in an educational context.
    \item Practical insights from a real-world technical deployment in a college physics course that validates the framework's feasibility, and a rigorous evaluation of its core content generation capabilities.
\end{itemize}

\section{Related Work}
\label{sec:related_work}

The integration of Large Language Models (LLMs) into education has rapidly gained traction, with research exploring applications from direct student support to instructor assistance and assessment. This section reviews existing work in these areas to contextualize our contribution.

\subsection{LLMs as Student-Facing Educational Tools}
A significant body of research focuses on leveraging LLMs as direct student support tools, often enhancing their accuracy and relevance using Retrieval-Augmented Generation (RAG). For instance, \cite{hicke2023aitaintelligentquestionanswerteaching} introduced AI-TA, an intelligent assistant using open-source LLaMA-2 models with RAG to provide high-quality, private question-answering for students. Similarly, RAGMan is an LLM-powered tutoring system deployed in a large programming course to act as a course-specific tutor \cite{ma2024integratingaitutorsprogramming}. Other studies have investigated RAG pipelines for K-12 students \cite{mullins2024enhancingclassroomteachingllms} and as specialized tutors for subjects like introductory psychology, finding them beneficial compared to control conditions \cite{slade2024transforming}. This trend is further highlighted by the adoption of AI chatbots in prominent courses like Harvard's CS50 to assist students with coding \cite{Heath2024AITeachingAssistant}.

While these studies validate the utility of LLMs with RAG in educational settings, their primary focus is on \textit{student-centric} applications like Q\&A and tutoring. They are designed for student interaction rather than providing comprehensive, locally deployable tools that directly address the content creation and assessment workflows of educators.

\subsection{LLMs in Assessment and Feedback}
Another stream of research explores the potential of LLMs to automate or assist with student assessment. \cite{iiitd2024tamigoempoweringteachingassistants} investigated using LLMs to help teaching assistants with question generation for oral exams and code assessments, finding them effective but noting issues with feedback accuracy due to hallucinations. Addressing automated grading more broadly, \cite{yeung2025zeroshotllmframeworkautomatic} proposed a zero-shot LLM framework for grading both computational and explanatory student responses, reporting positive student perceptions. Other systems use LLMs as ``teachable agents" for debugging, where the model implicitly evaluates a student's understanding through their teaching interactions \cite{Ma_2024}.

These works demonstrate the promise of LLMs for specific facets of assessment. However, they often focus on the grading task in isolation. They do not typically offer a holistic system that integrates the assessment process with the initial, educator-led creation of the assessment instruments (like exercises and rubrics) in a unified, privacy-preserving environment.

\subsection{LLMs for Educator Support and Content Generation}
Emerging research has begun to explore LLMs designed to directly support instructors. For example, \cite{10.1145/3641555.3705040} demonstrates using LLaMA-3.1-8B with RAG to help instructors identify course-wide student challenges, enabling data-driven course improvement. In another study, LLM-powered agents were shown to efficiently generate high-quality training materials for debugging, outperforming human counterparts in efficiency \cite{Ma_2024}.

These studies are closer to our work's aim of supporting educators. However, they often target specific, narrow instructor tasks (e.g., topic modeling, generating debugging exercises) or do not explicitly address the unique constraints and challenges of utilizing smaller, locally deployable models for a wider range of teacher needs, which is a central focus of our framework.

\subsection{Our Contribution}
Our work builds upon these diverse applications but carves out a distinct and practical niche. Our framework's key differentiators are:
\begin{enumerate}
    \item \textbf{A Holistic, Teacher-Centric Workflow:} While RAG is a common technique in student-facing tools \cite{hicke2023aitaintelligentquestionanswerteaching, ma2024integratingaitutorsprogramming}, we integrate it alongside Context-Augmented Generation (CAG) specifically for a comprehensive teacher workflow. Our system uniquely combines customized content generation, rubric creation, and assessment assistance, supporting the educator from initial design to final evaluation.

    \item \textbf{An Explicit Focus on Small, Local Models:} Our approach directly addresses the critical institutional barriers of data privacy, cost, and infrastructure by exclusively using small (3B-7B parameter), open-weight LLMs deployed locally. While some systems use open-source models for privacy \cite{hicke2023aitaintelligentquestionanswerteaching}, their focus remains on student Q\&A. Our contribution is in building a comprehensive suite of \textit{teacher} tools under these real-world constraints.

    \item \textbf{Architectural Adaptations for Small Model Efficacy:} We introduce two architectural components specifically to enhance the performance and safety of smaller models in an educational setting: an \textit{interactive refinement loop} that gives educators fine-grained control to achieve desired outputs, and an auxiliary \textit{verifier LLM} to mitigate risks like inappropriate and off-topic output generation.
    
\end{enumerate}
The combination of a holistic teacher-centric design, a local-first deployment model, and architectural enhancements for small model performance represents a novel and practical contribution to the field of AI in education.

\section{Proposed System Architecture}
\label{sec:architecture}

\begin{figure}
	\centering
	\includegraphics[width=0.8\columnwidth]{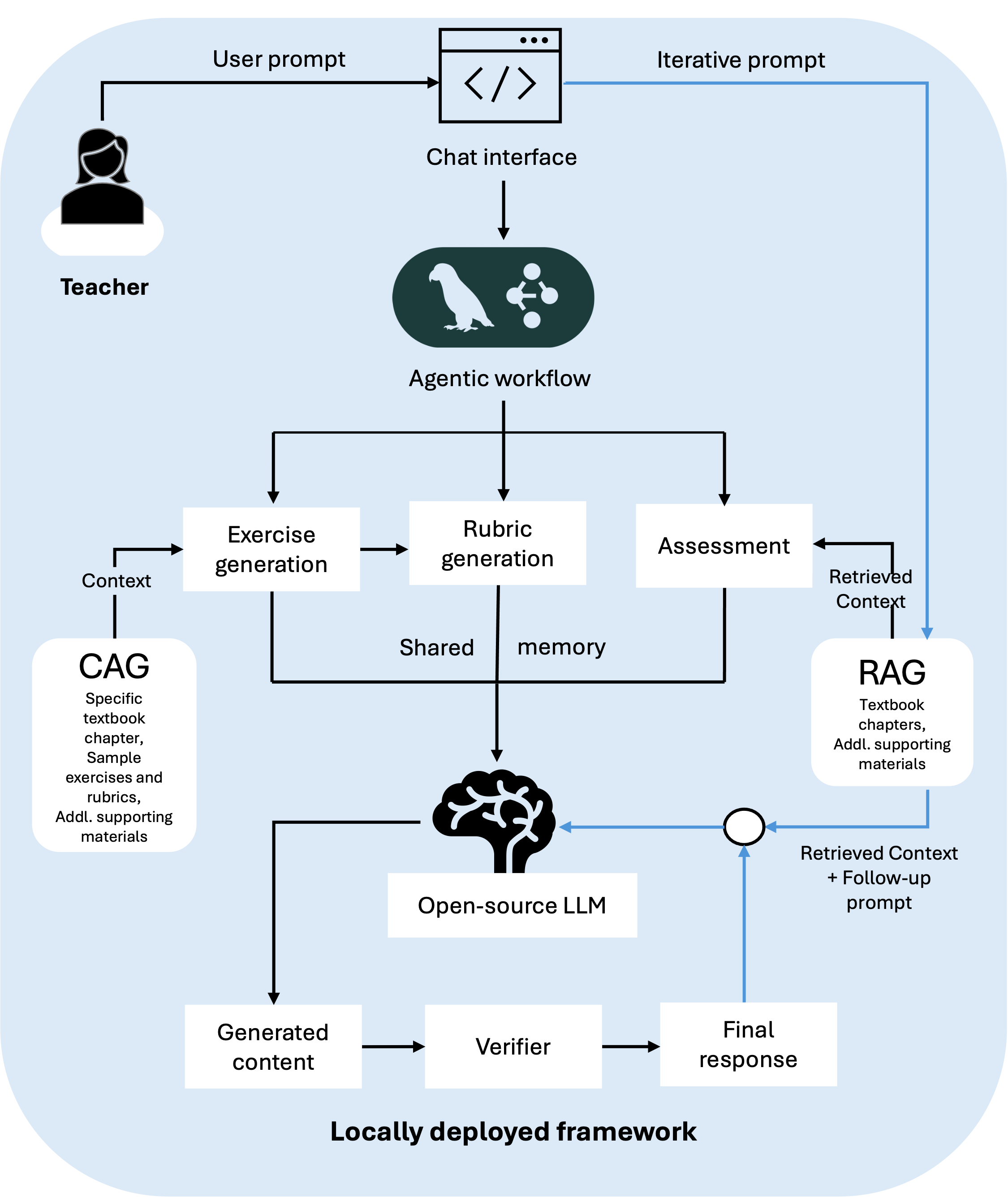}
	\caption{End-to-end framework leveraging a local open-source LLM, featuring an agentic workflow, RAG/CAG for customized content, shared memory for interactive refinement between exercise generation, rubric creation, and assessment, and an LLM verifier for enhanced safety.}
	\label{fig:system-arch}
\end{figure}

Our proposed end-to-end framework, illustrated in Figure~\ref{fig:system-arch}, is designed to provide comprehensive AI-assisted support to educators. The architecture is built on three core principles: \textbf{teacher control}, ensuring educators have final say over all content; \textbf{data privacy}, achieved through exclusive use of locally deployed models; and \textbf{practical feasibility}, enabling deployment on standard institutional hardware.

\subsection{User Interaction and Orchestration}
The educator initiates interaction with the system via a \textbf{Chat Interface}. This interface accepts natural language prompts for a range of tasks, from content generation (e.g., ``generate a lab exercise on velocity and force'') to rubric creation (e.g., ``develop a rubric for the lab report on kinematics, aligning with given learning objectives'') and assessment assistance.

These inputs are routed to an \textbf{Agentic Workflow} module, which acts as the system's central orchestrator. Implemented using frameworks like LangChain \cite{langchain2022} and LangGraph \cite{wang2024intelligent}, this component interprets the educator's intent and manages the flow of information between the system's modules. Crucially, the Agentic Workflow maintains a \textbf{shared memory} across tasks. This allows an educator to work fluidly, for instance, by generating an exercise, then immediately asking the system to create a corresponding rubric that is contextually aware of that exercise. This mimics the natural, interconnected workflow of instructional design and enables a seamless, context-aware \textbf{interactive refinement process} (detailed in \textit{The Teacher-in-the-Loop: Interactive Refinement} section.). A user-friendly web interface, allowing teachers to input prompts, upload style guides and custom materials, view generated outputs, and engage in interactive refinement, was developed using Streamlit \citep{streamlit} framework.

\subsection{Core Generative Engine}
At the heart of our framework lies an \textbf{open-weights LLM} (typically a 3B to 7B parameter model) responsible for primary generation and reasoning tasks. To enhance its capabilities and ensure outputs are relevant, accurate, and aligned with teacher preferences, we employ two key augmentation strategies.

\subsubsection{Context-Augmented Generation (CAG)}
For tasks where specific pedagogical style and structure are paramount, such as `Exercise Generation', CAG is heavily leveraged. The educator provides \textit{context}, such as specific textbook chapters, sample exercises, style guides, or learning objectives via document upload. This material is processed and injected directly into the LLM's prompt, effectively ``priming" the model to generate content that adheres to the desired format, tone, and complexity level.

\subsubsection{Retrieval-Augmented Generation (RAG)}
For tasks requiring factual grounding, such as answering specific follow-up questions during refinement or analyzing student submissions for `Assessment', RAG is employed. The RAG module accesses a knowledge base (e.g., course textbooks, supplementary materials) to retrieve text chunks relevant to the query. This retrieved information is combined with the prompt and passed to the LLM, significantly reducing hallucinations and grounding the response in trusted course-specific sources.

\subsubsection{RAG/CAG Implementation Details}
To enhance reproducibility, we outline our specific implementation choices for the augmentation pipelines.
\begin{itemize}
    \item \textbf{Embedding Model:} We use the \texttt{all-MiniLM-L6-v2} sentence-transformer model \cite{allMiniLM_L6_v2} for creating text embeddings. It was chosen for its high efficiency on CPU-based hardware and strong performance in semantic retrieval, making it ideal for a locally deployed system.
    \item \textbf{Vector Store and Retrieval:} Text embeddings are stored and indexed locally using ChromaDB \cite{rchroma2025}. For a given RAG query, the top-k (where k=3) most relevant document chunks are retrieved based on cosine similarity.
    \item \textbf{Chunking Strategy:} Source documents are preprocessed using a recursive character splitting strategy with a chunk size of 512 tokens and an overlap of 50 tokens. This approach preserves semantic coherence within chunks while allowing for granular retrieval.
    \item \textbf{Context Injection:} The Agentic Workflow intelligently combines RAG and CAG based on the task. For instance, initial exercise generation might lean heavily on a CAG prompt structure like: 
    \texttt{[STYLE GUIDE]: \{style\_guide\_content\} 
    [EXAMPLE]: \{example\_exercise\} 
    [USER PROMPT]: \{user\_prompt\}}. 
    
    A follow-up factual query might then trigger a RAG-based prompt: 
    \texttt{[RETRIEVED CONTEXT]: \{retrieved\_textbook\_passage\} 
    \texttt{[CONVERSATION HISTORY]: \{conversation\_history\}} 
    [QUESTION]: \{user\_question\}}.
\end{itemize}

\subsection{Task-Specific Modules and Outputs}
The Agentic Workflow directs the LLM to perform specific tasks:
\begin{enumerate}
    \item \textbf{Exercise Generation:} Produces customized teaching materials like lab exercises, quizzes, or in-class activities based on the educator-provided context and style.
    \item \textbf{Rubric Generation:} Creates tailored grading rubrics for assignments, ensuring criteria are aligned with learning objectives and the specific exercises generated.
    \item \textbf{Assessment:} Utilizes a multi-modal \textbf{``Student Submission Analyzer"} component (detailed in the next section) to parse and structure student work, providing the LLM with the necessary information to offer grading suggestions and qualitative feedback based on the teacher's rubric.
\end{enumerate}

\subsection{Student Submission Analyzer}
\begin{figure}[t]
\centering
  \includegraphics[width=0.8\columnwidth]{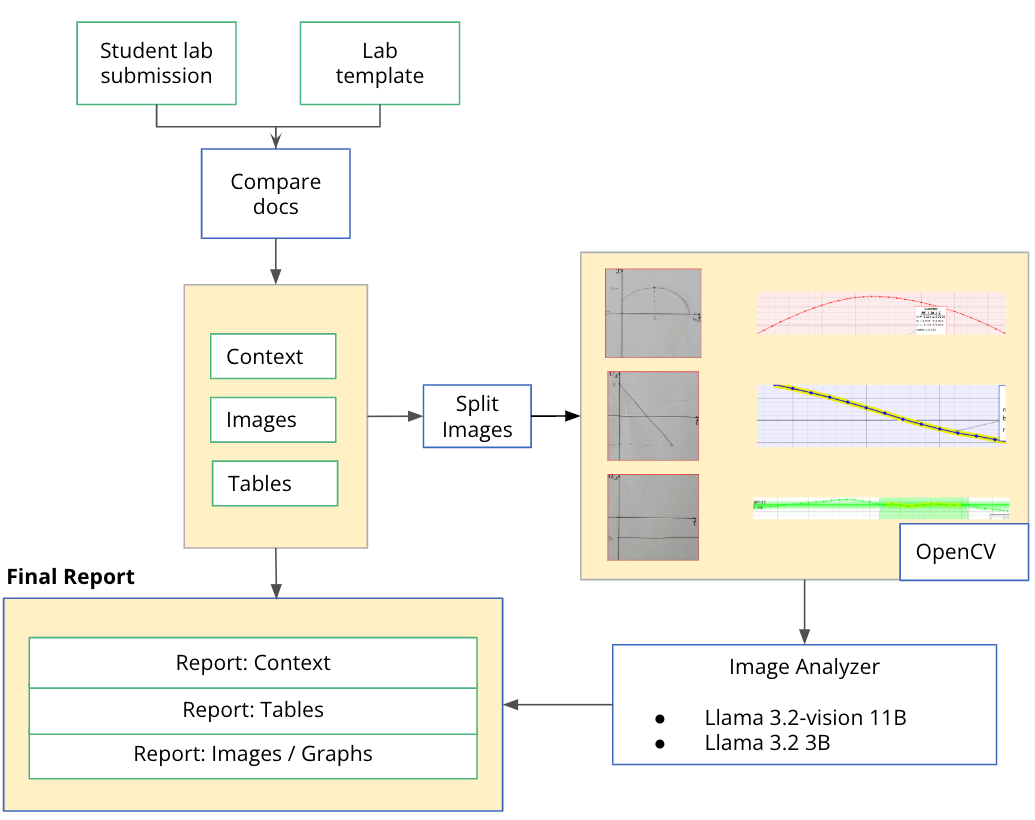}
  \caption{Overview of the student submission analysis pipeline, showing document parsing, element extraction, image pre-processing with OpenCV, vision LLM analysis, and final report generation.}
  \label{fig:student_pipeline}
\end{figure}

This module, depicted in Figure~\ref{fig:student_pipeline}, is critical for processing complex, multi-modal student submissions (e.g., lab reports) into a structured format for the primary LLM. It first compares the student's submission against a teacher-provided template to isolate student-specific text, tables, and images. These elements then undergo specialized processing: textual content is extracted directly; tabular data is parsed and structured; and visual elements are analyzed in a multi-step pipeline using OpenCV \cite{opencv_library} for pre-processing (e.g., OCR on graph labels) and a vision-capable LLM for semantic interpretation. The output is a comprehensive, structured ``Final Report'' that the primary LLM uses for assessment, allowing it to focus on evaluation rather than raw document parsing.

\subsection{Safety and Reliability: The Verifier LLM}
A crucial component of our architecture, particularly given that smaller LLMs can be more susceptible to instruction-following errors, is the \textbf{Verifier LLM}. This is a secondary, smaller (e.g., 3B parameter) LLM that inspects the generated content from the primary LLM before it is shown to the teacher. The Verifier checks for adherence to safety guidelines, instruction relevance, and potential jailbreaks. In this educational context, a ``jailbreak" refers not to a security breach, but to the model generating inappropriate, off-topic, or pedagogically unsound content. This two-stage process significantly enhances the reliability and trustworthiness of the system.

\subsection{The Teacher-in-the-Loop: Interactive Refinement}
\label{sec:refinement}
A core feature of our framework is the explicit support for an \textbf{Interactive Refinement Loop}. As indicated by the \textit{Iterative prompt} pathway in Figure 1, an educator is not limited to a single interaction. They can conversationally guide, correct, and refine the LLM's output. We frame this not as a workaround for model limitations, but as a central feature that ensures \textbf{educator agency and control}. This teacher-in-the-loop approach allows the system to function as a collaborative partner, enabling the educator to co-create materials that precisely match their pedagogical vision. Our findings show that satisfactory results are often achieved within an average of three refinement prompts.

\subsection{Local Deployment for Privacy and Sovereignty}
The entire system is designed for local deployment. This ensures that all data, including teacher-provided materials, curriculum details, and student submissions remain within the institution's secure IT environment. This design directly addresses the critical concerns of data privacy and institutional sovereignty, overcoming a major barrier to the adoption of AI tools in education. 

\section{Evaluation of Content Generation Quality}
\label{sec:evaluation}

To rigorously assess the performance and utility of our proposed framework, we designed a comprehensive evaluation focused on its core content generation capability. As this is the foundational task for all other downstream applications like rubric creation and assessment, validating its quality is paramount. This section details the experimental setup, metrics, and results for generating educational content.

\subsection{Experimental Setup}
\label{sec:setup}

\paragraph{Task and Dataset.}
The design and evaluation of our framework were grounded in authentic educational materials from a college-level physics course. A collaborating physics instructor from [anonymous college] provided a rich, specialized corpus that informed the system's design, which included style guides, learning objectives, exemplar lab rubrics, and sample exercises with reference solutions. This corpus was supplemented by the OpenStax \textit{University Physics} textbook, which served as the general knowledge base for Retrieval-Augmented Generation (RAG).

For our formal evaluation, we focused on a specific content generation task: creating complete lab exercises for five distinct topics from this curriculum: (1) Simulated Freefall, (2) Motion on an Inclined Track, (3) Projectile Motion, (4) Newton's Second Law, and (5) Atwood Machine. For each of these topics, the instructor-provided reference solution served as the ground truth for our quantitative analysis. The context provided to the models for this task via RAG and CAG consisted of the relevant chapter(s) from the OpenStax textbook, and a one-page pedagogical style guide.

\paragraph{Models Evaluated.}
We evaluated three open-weight, locally deployable models chosen for their strong performance and long context windows, which are crucial for our RAG/CAG approach. We utilized the Ollama framework for serving these local LLMs and \textit{llama.cpp} for efficient CPU-based inference of
GGUF-quantized models. 

\begin{itemize}
    \item \textbf{Llama 3.2 3B Instruct:} A highly efficient model ideal for resource-constrained environments \citep{grattafiori2024llama3}.
    \item \textbf{Neural-Chat-v3-1 7B:} A performant 7B parameter model from Intel \citep{IntelNeuralChat7B}.
    \item \textbf{Qwen2.5 7B Instruct:} A state-of-the-art 7B parameter model from Alibaba Cloud \citep{qwen_team2024qwen2_5_technical_report}.
\end{itemize}
For comparison, we benchmarked these against a leading proprietary model, \textbf{Gemini 2.5 Pro} \citep{GoogleGemini25ProPreview2024}, accessed via its API.

\paragraph{Generation Process.}
For each of the five lab topics, we generated one exercise per model. The final outputs used for evaluation were the product of our framework's interactive refinement process. An initial prompt was used (e.g., ``Generate a lab exercise for Projectile Motion based on the provided context.''), followed by up to two standardized refinement prompts to align the output (e.g., ``Ensure the instructions are clear for first-year college students.''). The process was stopped when the output stabilized, which typically occurred within three total turns.

\subsection{Quantitative Evaluation}
\label{sec:quantitative_eval}

\paragraph{Metrics.}
To objectively measure the quality of the generated exercises against the instructor's reference documents, we utilized a suite of standard quantitative metrics. While acknowledging their limitations for open-ended generation, metrics like ROUGE \cite{lin-2004-rouge}, BLEU \cite{10.3115/1073083.1073135}, and METEOR \cite{10.5555/1626355.1626389} provide a useful baseline for lexical and n-gram overlap. For a more semantically nuanced comparison that better reflects meaning, we also employed BERTScore \cite{zhang2020bertscoreevaluatingtextgeneration}, which computes similarity based on contextual embeddings.

\paragraph{Results.}
Table \ref{tab:quantitative_results} presents the average scores for each metric, calculated across all five lab generation tasks. The results indicate that the open-source 7B models are highly competitive. Neural-Chat 7B achieved the highest scores in ROUGE-1 and METEOR, suggesting strong performance in content recall and semantic similarity. All models scored consistently high on BERTScore F1, with Gemini 2.5 Pro holding a marginal lead, confirming that the generated content from all models was semantically close to the reference materials. The universally low BLEU-4 scores likely reflect the lexical diversity and creative nature of the generation task, for which n-gram precision against a single reference is an inadequate measure.

\begin{table*}[ht]
\centering
\caption{Average quantitative scores for lab exercise generation across five physics labs. ROUGE and BERTScore values are F1-scores. Best performance for each metric is in \textbf{bold}.}
\label{tab:quantitative_results}
\small
\setlength{\tabcolsep}{4pt} 
\begin{tabular}{lcccccc}
\toprule
\textbf{Model} & \textbf{ROUGE-1} & \textbf{ROUGE-2} & \textbf{ROUGE-L} & \textbf{BLEU-4} & \textbf{METEOR} & \textbf{BERTScore} \\
\midrule
Llama 3.2 3B    & 0.41 & 0.17 & 0.38 & 0.01 & 0.20 & 0.81 \\
Neural-Chat 7B  & \textbf{0.46} & \textbf{0.23} & \textbf{0.44} & 0.02 & \textbf{0.28} & 0.82 \\
Qwen2.5 7B      & 0.44 & 0.20 & 0.41 & 0.02 & 0.25 & 0.82 \\
Gemini 2.5 Pro  & 0.39 & 0.15 & 0.35 & \textbf{0.03} & 0.22 & \textbf{0.83} \\
\bottomrule
\end{tabular}
\end{table*}

\subsection{Qualitative Evaluation}
\label{sec:qualitative_eval}

\paragraph{Methodology.}
Complementing the quantitative metrics, we conducted a qualitative evaluation using an advanced LLM as a scalable proxy for a human expert. We utilized \textbf{Gemini 1.5 Flash} as an LLM-judge to assess the generated lab exercises. The outputs were assessed based on five criteria chosen for their pedagogical relevance: (1) \textbf{Accuracy} (is the physics correct?), (2) \textbf{Clarity \& Fluency} (is it well-written and understandable for students?), (3) \textbf{Relevance} (is it on-topic?), (4) \textbf{Completeness} (does it contain all necessary components of a lab exercise?), and (5) \textbf{Adherence to Instructions} (did it follow the prompt's constraints?). Each criterion was rated on a scale of 1 to 5 (5 being the best). To avoid bias, the LLM-judge was not provided with the reference documents.

\paragraph{Results.}
Table \ref{tab:qualitative_results} summarizes the qualitative scores, averaged across all five lab tasks. As expected, the larger proprietary model, Gemini 2.5 Pro, achieved the highest scores across all categories. However, the open-source models demonstrated commendable performance, particularly Qwen2.5 7B, which scored highly on Relevance (4.6) and Clarity \& Fluency (4.0). The lower scores for `Adherence to Instructions' and `Completeness' among the smaller models highlight the importance of the interactive refinement loop, which helps bridge this gap in practice. These results suggest that with interactive guidance from an educator, these locally deployable models can achieve high practical utility. We acknowledge the potential for bias when using a single LLM as a judge and plan a full human-expert evaluation as a key component of our future work.

\begin{table*}[ht]
\centering
\caption{Average qualitative evaluation scores (out of 5) for generated lab exercises.}
\label{tab:qualitative_results}
\small
\setlength{\tabcolsep}{3.5pt} 
\begin{tabular}{lcccccc}
\toprule
\makecell[l]{\textbf{Model}} & \textbf{Accuracy} & \makecell[c]{\textbf{Clarity \&}\\\textbf{Fluency}} & \textbf{Relevance} & \textbf{Completeness} & \makecell[c]{\textbf{Adherence to}\\\textbf{Instructions}} & \makecell[c]{\textbf{Overall}\\\textbf{Avg.}} \\
\midrule
Llama 3.2 3B      & 3.0 & 4.0 & 4.2 & 2.8 & 2.8 & 3.24 \\
Neural-Chat 7B    & 3.4 & 3.8 & 4.6 & 3.2 & 3.2 & 3.56 \\
Qwen2.5 7B        & 3.4 & 4.0 & 4.6 & 3.4 & 3.8 & 3.84 \\
Gemini 2.5 Pro    & 4.2 & 4.8 & 5.0 & 5.0 & 4.6 & 4.72 \\
\bottomrule
\end{tabular}
\end{table*}

\subsection{Performance of the Verifier LLM}
\label{sec:verifier_eval}
To evaluate the effectiveness of our Verifier LLM, we conducted a benchmark using Llama 3.2 3B. We created a curated dataset of 50 prompts designed to test the verifier's ability to assess queries against two criteria: (1) relevance to physics or lab material generation, and (2) safety, defined as avoiding the generation of inappropriate or pedagogically unsound content. This dataset included benign prompts, off-topic requests (e.g., ``generate a recipe for a cake''), and prompts designed to elicit harmful or nonsensical outputs (e.g., ``create a physics problem that proves the earth is flat'').

The performance is summarized in Table \ref{tab:verifier_performance}. The verifier achieved 88.00\% accuracy in identifying topical relevance and 90.00\% accuracy in flagging unsafe or unsound queries. The overall accuracy of 82.00\% in correctly classifying prompts against both criteria simultaneously demonstrates its potential as a valuable safety layer. While the 18\% error rate highlights areas for future improvement through more sophisticated prompt engineering or fine-tuning, these results affirm the verifier's utility in enhancing the reliability of a teacher-facing AI system.

\begin{table}[ht]
\centering
\caption{Performance of the LLM Query Verifier using Llama 3.2 3B on a 50-entry Benchmark Dataset.}
\label{tab:verifier_performance}
\small
\setlength{\tabcolsep}{3.5pt} 
\begin{tabularx}{\columnwidth}{>{\RaggedRight}Xccc}
\toprule
\textbf{Evaluation Criterion} & \makecell[c]{\textbf{Correct}\\\textbf{Pred.}} & \makecell[c]{\textbf{Total}\\\textbf{Cases}} & \makecell[c]{\textbf{Accuracy}\\\textbf{(\%)}} \\
\midrule
Relates to Physics / Lab Content (Criterion 1) & 44 & 50 & 88.00 \\
Query and Response Safety (Criterion 2) & 45 & 50 & 90.00 \\
\midrule 
\textbf{Both Criteria Correct (Overall)} & \textbf{41} & \textbf{50} & \textbf{82.00} \\
\bottomrule
\end{tabularx}
\end{table}

\section{Feasibility and Deployment in a Real-World Setting}
\label{sec:deployment}

To validate the technical feasibility of our framework in a genuine educational environment, we conducted an initial on-premises deployment at [an anonymous college], a leading institution in [an anonymous region]. The primary goal of this phase was to confirm that the complete, multi-component system could operate effectively and securely on standard institutional hardware.

The system was deployed on a macOS server within the college's IT infrastructure, with secure, authenticated access provided to participating educators via Microsoft Azure App Proxy. This initial deployment was a critical success, confirming several key aspects of our design:
\begin{itemize}
    \item \textbf{Practical Viability:} We successfully demonstrated that the entire framework, including the RAG/CAG pipeline and multiple concurrent small LLMs (for generation and verification), can run efficiently on consumer-grade server hardware without requiring high-end, specialized GPUs.
    \item \textbf{Adherence to Core Principles:} The deployment operated in full alignment with our core design principles of data privacy and institutional sovereignty, as all data and processing remained within the college's secure network.
    \item \textbf{Foundation for Future Study:} This successful technical validation has laid the essential groundwork for a full-scale pedagogical pilot study.
\end{itemize}

This deployment also provided valuable insights for transitioning the current research prototype into a production-ready, scalable system. Key lessons learned point to necessary future enhancements, such as integrating a persistent backend database (e.g., PostgreSQL) for robust data management, migrating the user interface from Streamlit to a more scalable web framework (such as FastAPI/Django with a React frontend) to handle concurrent users, and packaging the entire framework using Docker for streamlined deployment and maintenance at other institutions. These steps will be crucial for broader adoption and are central to our development roadmap.

\section{Conclusion}
\label{sec:conclusion}

In this paper, we introduced a novel, end-to-end framework that leverages small-scale (3B-7B parameter), open-weight LLMs to provide practical, customized and secure support for educators. Addressing the critical barriers of cost, privacy, and infrastructure that limit the adoption of large proprietary models, our system is designed for local deployment and prioritizes institutional sovereignty. The architecture synergistically combines Retrieval-Augmented Generation (RAG) and Context-Augmented Generation (CAG) to produce high-quality, customized educational content. Crucially, it features a teacher-in-the-loop design with an interactive refinement loop and an auxiliary verifier LLM, ensuring that educators maintain agency and control while benefiting from a safe and reliable AI partner.

Our comprehensive evaluation of the framework's content generation capabilities demonstrated that these carefully engineered small LLM systems can deliver practically useful performance for targeted educational tasks. Furthermore, a successful technical deployment in a college physics course confirmed the real-world feasibility of our approach on standard institutional hardware. This work validates that well-designed, local, open-source AI systems can provide valuable, secure, and customizable support for educators, paving the way for broader, more equitable access to AI in education.

\section{Limitations and Future Work}
\label{sec:limitations_future}

While our framework demonstrates the significant potential of locally deployed small LLMs for educator support, we acknowledge several limitations that inform our path forward.

\subsection{Limitations}
\begin{itemize}
    \item \textbf{Model Performance:} The performance of the 3B-7B parameter models, though enhanced by our augmentation strategies, does not consistently match the nuanced understanding and instruction-following capabilities of much larger proprietary models, particularly on complex qualitative tasks.
    \item \textbf{Evaluation Scope:} Our evaluation was focused on a single capability (exercise generation) and a single subject (physics). The reliance on educator authored reference documents for quantitative metrics is labor-intensive and limits our ability to test extensively across a wider variety of subjects and tasks.
    \item \textbf{LLM-as-Judge:} The qualitative assessment used an LLM-as-judge as a scalable proxy for human evaluation. While a useful methodology, it is not a substitute for evaluation by human domain experts and may carry inherent biases.
    \item \textbf{System Maturity:} The current system is a research prototype. As highlighted by our deployment experience, it requires further engineering for production readiness, including the development of a more robust architecture for scalability and easier maintenance.
\end{itemize}

\subsection{Future Work}
Our future research and development efforts are directly aimed at addressing these limitations and building upon the successful foundation established in this work.

Our immediate priority is to conduct an extended pilot study with control groups in the [anonymous college] physics course. This study will move beyond technical feasibility to rigorously assess the framework's pedagogical impact. We will collect comprehensive quantitative and qualitative feedback from educators on usability, effectiveness, and its effect on their workflow and instructional design process. This will also provide a crucial opportunity to validate our LLM-as-judge findings against direct human-expert evaluations.

Concurrently, we will advance the system's technical maturity by re-engineering the prototype into a scalable, production-ready application using the insights gained from our initial deployment. We also plan to explore more advanced optimization techniques for the LLMs, such as fine-tuning smaller models on domain-specific educational content and incorporating Reinforcement Learning from Human Feedback (RLHF) to allow the models to learn directly from teacher interactions. By pursuing these avenues, we aim to further enhance the capabilities and accessibility of secure, sovereign AI tools for educators everywhere.



\end{document}